\documentclass{nature}

\usepackage{graphicx}
\makeatletter
\let\saved@includegraphics\includegraphics
\AtBeginDocument{\let\includegraphics\saved@includegraphics}
\renewenvironment*{figure}{\@float{figure}}{\end@float}
\makeatother
\usepackage{subfigure}

\title{Toward quantitative fractography using convolutional neural networks}

\author{Stylianos Tsopanidis$^{1}$, Raúl Herrero Moreno$^2$ \& Shmuel Osovski$^3$}

\begin{document}
\def\finp{\\ \hspace*{0.5cm}}
\def\r {{\bf r}}
\def\sc {\hat{\sigma}}
\def\tT {\delta T}
\def \tD {\delta D}
\def \tF {\delta F}
\def \tx {\delta x}
\def\tsig {\delta \sigma}
\def \ps {\delta s}
\def \tp {\delta p}
\def\eps {\epsilon^{'}}
\def\bsigma {\boldsymbol{\sigma}}
\def\btd{\boldsymbol{\dot{T}}}
\def\bt{\boldsymbol{T}}
\def\bl{\boldsymbol{L}}
\def\msr {  \underline{\underline{D}}}
\def\ml {  \underline{\underline{L}}}
\def\mdl {  \underline{\underline{\dot{L}}}}
\def\mo{  \underline{\underline{\Omega}}}
\def\mdsr {  \underline{\underline{\dot{D}}}}
\def\sr { \underline{\underline{d}}}
\def\ts {\underline{\underline{\sigma}}}
\def\tsm {\underline{\underline{\Sigma}}}
\def\smij {\Sigma_{ij}}
\def\smik {\Sigma_{ik}}
\def\sij {\sigma_{ij}}
\def\avsij {< \sij >}
\def\mc{\underline{\underline{C}}}
\def\sik {\sigma_{ik}}
\def\uvol {\frac{1}{\vert \Omega \vert}}
\def\dij {d_{ij}}
\def\do {\partial \Omega}
\def\into {\int_{\Omega}}
\def\intdo {\int_{\partial \Omega}}
\def\dmij {D_{ij}}
\def\dmik {D_{ik}}
\def\ux {\underline{x}}
\def\uv {\underline{v}}
\def\work {< \ts : \delta \sr >}
\def\works { < \ts : \sr >}
\def\bibsection{\section*{References}}
\maketitle

\begin{affiliations}
 \item Department of Continuum Mechanics and Structural Analysis. University Carlos III of Madrid. Avda. de la Universidad, 30. 28911 Legan{\'e}s, Madrid, Spain.
 \item Sertec, AEROSERTEC Group, Avda. Rita Levi Monatalcini, 14, 28906 Getafe, Madrid, Spain.
 \item Faculty of Mechanical Engineering, Technion - Israel Institute of Technology, Haifa, Israel.
 
 Correspondence: Shmuel Osovski (shmuliko@technion.ac.il)
\end{affiliations}
\pagebreak 
\begin{abstract}
The science of fractography revolves around the correlation between topographic characteristics of the fracture surface and the mechanisms and external conditions leading to their creation. While being a topic of investigation for centuries, it has remained mostly qualitative to date. A quantitative analysis of fracture surfaces is of prime interest for both the scientific community and the industrial sector, bearing the potential for improved understanding on the mechanisms controlling the fracture process and at the same time assessing  the reliability of computational models currently being used for material design.  With new advances in the field of image analysis, and specifically with machine learning tools becoming more accessible and reliable, it is now feasible to automate the process of extracting meaningful information from fracture surface images. Here, we propose a method of identifying and quantifying the relative appearance of intergranular and transgranular fracture events from scanning electron microscope images. The newly proposed method is based on a convolutional neural network algorithm for semantic segmentation. The proposed method is extensively tested and evaluated against two ceramic material systems ($Al_2O_3$,$MgAl_2O_4$) and shows high prediction accuracy, despite being trained on only one material system ($MgAl_2O_4$). While here attention is focused on brittle fracture characteristics, the method can be easily extended to account for other fracture morphologies, such as dimples, fatigue striations, etc. 
\end{abstract}

The fracture process of materials is governed by both extrinsic (e.g. imposed loading, environmental conditions) and intrinsic (microstructure) characteristics. One may thus expect, that the fracture surface will contain evidence regarding the influence of both the intrinsic and extrinsic characteristics of the fracture process. Fractography is a powerful tool employed to study fracture surfaces and extract information regarding material properties and factors leading to failure. The main objective of fractography is the topographic characterization of the fractured surfaces, aiming to identify and classify the different operating fracture mechanisms, and correlating them with the material's microstructure, its mechanical behavior, and the conditions leading to failure. While the study of fracture surfaces dates back to the sixteenth century \cite{ASM}, the lack of quantitative robust methods to describe the complex geometries which compose the fracture surface, has rendered these studies to be mostly qualitative in nature. 

Currently, the quantitative and qualitative examination of the fracture surfaces is a cumbersome process, performed manually by experienced personnel,thus demanding substantial labor and qualifications. Moreover, the heavy reliance on the human factor in the process is prone to errors in quantitative estimations. The automation of this process  and the quantitative study of fracture surface has posed a long standing challenge. In terms of industrial relevance, the strength of quantitative, automated fractography lies mostly in the ease and reliability of finding the root cause of fracture in failure analysis of components, a process which today is only performed by very experienced individuals. In terms of scientific value, it holds the promise of unraveling new and exciting insights as to the way materials fail. It is sufficient to mention just some of the lessons learned through careful fractographic analysis to realize the importance of making new progress in the field. Among those are the micro-mechanisms leading to ductile fracture \cite{tipper}, the process of cleavage in metals \cite{pineau1} and the origin of striations in fatigue fracture \cite{pineau2}. 

In the context of brittle fracture of ceramic materials, two primary modes of crack propagation are often observed: transgranular and intergranular. The former mode (transgranular) results from crack propagation through a cleavage mechanism of the grains, while in the later (intergranular) the crack propagation is facilitated through grain separation along grain boundaries. The two propagation modes provide evidence as to the grain boundaries strength, environmental conditions and material anisotropy. In many scenarios, both crack growth mechanisms will be activated and their relative occurrence can testify as to the applied loading rate \cite{marbel,hu2012mechanisms}, chemical environment and loading history \cite{keren}, as well as to the presence of changes in the initial microstructure \cite{sigma}.

With recent progress in the fields of materials design, grain boundary engineering and computational modeling of fracture, it is now feasible to find an optimal microstructure along with establishing the processing routes to manufacture it. In the field of brittle fracture, the role of grain boundary properties on transgranular fracture and the resulting fracture toughness was recently addressed \cite{molinari} using finite elements calculations, leading to the conclusion that microstructural engineering with respect to grain morphology can lead to enhanced performance. Similar observations were made for metastable $\beta$ Ti alloys where the fracture process is predominantly along grain boundaries \cite{soijf}. The use of computational models for microstructural design, requires that they can be validated against experiments. While crack growth resistance curves can be reproduced in simulations and compared with experimental data, combining those with quantitative data extracted directly from experimentally obtained fracture surfaces will tremendously increase the reliability of such models. One such comparison, for example, is the relative area fractions of transgranular and intergranular crack propagation modes obtained from experiments and numerical simulations\cite{molinari}   

Some of the initial approaches toward the automation of fractographic features extraction utilized computer vision and image processing techniques to classify and characterize the optical microscopy or the Scanning Electron Microscopy (SEM) fracture images. Hu et. al.\cite{Hu2017} employed edge detection and peak finding algorithms to determine the fatigue crack growth from optical microscope fracture images, while in the work of Kosarevych\cite{Kosarevych2013} the histogram of brightness information enabled the feature segmentation of SEM fracture images. Similarly,  various texture analysis methods (gray level co-occurrence matrix, run length statistical analysis, box-counting, Fourier power spectrum, etc.) were utilized to identify different fracture surface morphologies and characterizing micrographs or fractographs of several steels\cite{Kenjiro1993,Dutta2014}. Focusing our attention back to brittle fracture mechanisms, Yang et. al\cite{yang1991sem} proposed an automated procedure for the quantification of transgranular vs intergranular fracture from SEM images. The method proposed by Yang et. al\cite{yang1991sem} is based on analyzing local intensity profiles obtained by SEM secondary-electron detector. The profiles are then analyzed with respect to the average grain size and following the procedure detailed in \cite{yang1991sem}, regions are determined as transgranular or intergranular. While offering an automated method for quantification of brittle fracture surfaces, this method still requires additional inputs from the user which reduces its robustness. Similarly, the application of this method requires the user to follow several restricting assumptions regarding the size distribution of microstructural features, and will pose a challenge for highly irregular surfaces, where achieving the required contrast simultaneously with a larger depth of focus becomes problematic for images taken over large areas.  

More recently, advances in machine learning and artificial intelligence have captured the attention of the materials characterization community as a way to classify and quantify data available from various characterization techniques. Microstructure classification and defect analysis have been shown to be feasible through the usage of machine learning methods\cite{Chowdhury2016,Zhang2019,Li2018,Gola2018}. In the field of fractography, a few attempts to use these methods to develop computerized models for quantitative fractography have been published. More specifically, Bastidas-Rodriguez et. al\cite{Bastidas-Rodriguez2016} combined texture analysis techniques with non-linear machine learning classifiers (artificial neural networks and support vector machine) in order to classify optical microscope fracture images in three different modes: \textit{brittle sudden}, \textit{ductile sudden} and \textit{fatigue}. Additionally, in a recent work \cite{Konovalenko2018} convolutional neural networks have been used for the detection of dimples and edges on SEM ductile fracture images of titanium alloys. To the best of our knowledge, this is the only published work using exclusively machine learning methods for classification of fracture surfaces.    

\begin{figure}
	\centering
	\includegraphics[width=140mm]{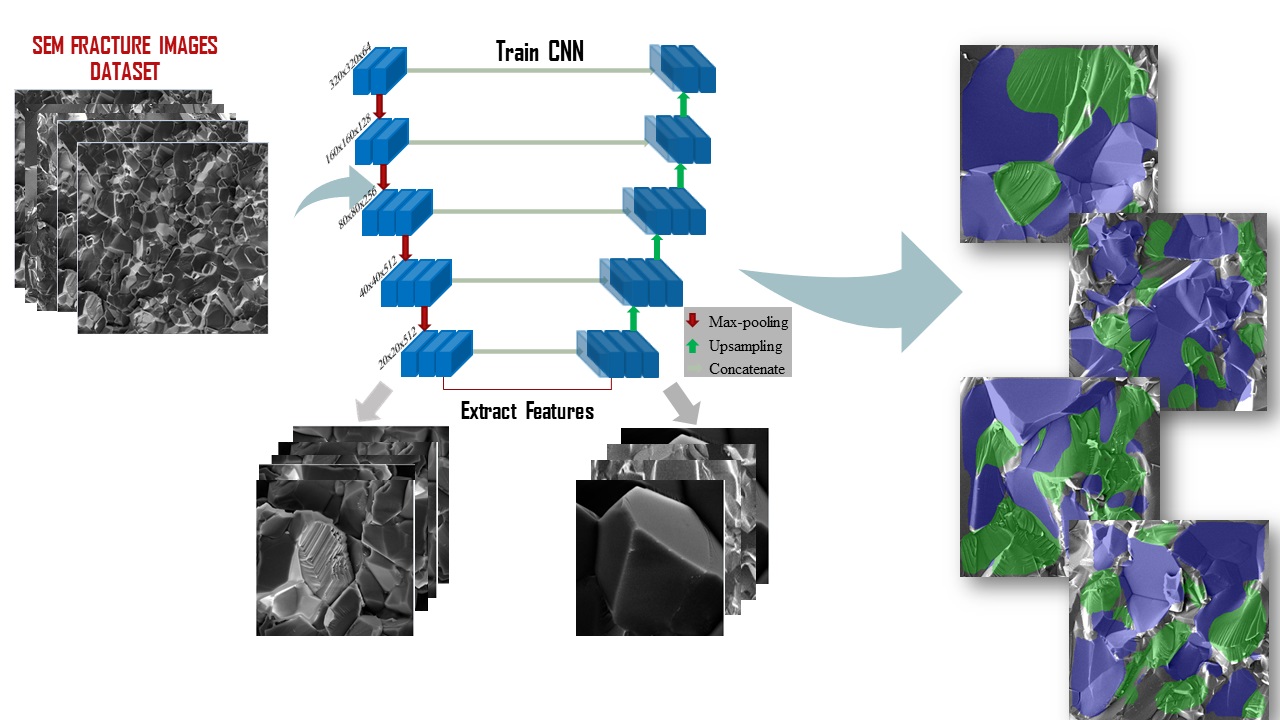}
	\caption{Schematic flow chart of the method used for the classification of fracture surfaces as intergranular and transgranular .}
	\label{chart}
\end{figure}

Here, we present a new method for the automatic topographic characterization of fracture surfaces, based on Convolutional Neural Network (CNN) Semantic Segmentation\cite{chen2014semantic}. The architecture of modern deep learning algorithms for semantic segmentation is divided into two main parts: the encoder and the decoder part. The encoder is usually the backbone architecture of some of the most efficient convolutional neural networks classifiers, such as VGG \cite{Simonyan2014}, ResNet \cite{He2016} or GoogLeNet \cite{Szegedy2015}. Using this architecture for the encoder, besides the obvious benefit of using a well-proven network, allows us to make use of transfer learning \cite{Donahue2013,Zeiler13}. As a result of previous research work, one can find pre-trained weights for each one of the above mentioned encoder architectures on different datasets \cite{coco,Imagenet}, and by fine-tuning\cite{Zeiler13} them it is possible to achieve high accuracy training for a new dataset. This method significantly reduces the computation time for training and performs considerably better when compared to random weights initialization. This is extremely useful when training networks for a relatively small training dataset. The CNN, after being trained on SEM fracture images of brittle material, is able to classify every pixel of any new SEM image of the same material. During the training process the network is using a dataset consisted of SEM images of fracture surfaces of Magnesium Aluminate Spinel ($MgAl_2O_4$) samples, in order to learn how to extract characteristic features of the different fracture modes that will allow it to perform accurate predictions on any new image; this function is graphically demonstrated in Figure \ref{chart} .  

\section*{Results and discussion}

\begin{figure}
	\centering
	\subfigure[]{\label{sem_image}\includegraphics[width=70mm,height=70mm]{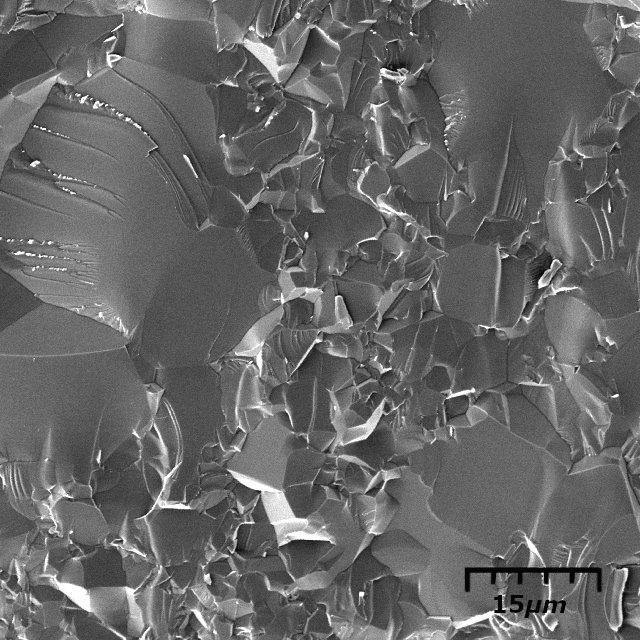}}
	\hspace{1mm}
	\subfigure[]{\label{sem_annot}\includegraphics[width=70mm,height=70mm]{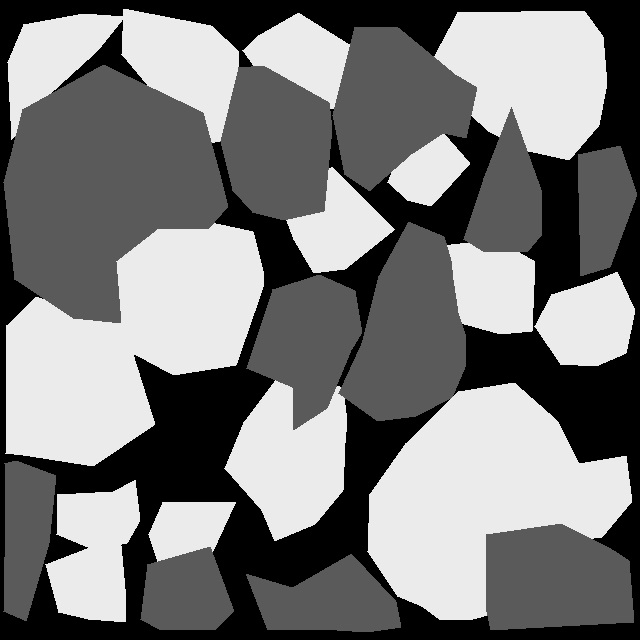}}
	\caption{Original (a) and annotated (b) SEM image used in the CNN training. The white areas in the annotated image are \textit{intergranular} modes, gray are \textit{transgranular} and the black non-annotated pixels are considered as the \textit{background}.}
	\label{sem_example_image}
\end{figure}

The aim of quantitative fractography is to find well defined mathematical descriptors which represent the complex topographic morphology of fracture surfaces. These complex morphologies encode the interlacing of the material's microstructure and intrinsic mechanical properties with the externally applied loads and environmental conditions\cite{barak2019}. For brittle materials, such as ceramics, the fracture process is often observed to take place using two alternating micromechanisms: transgranular and intergranular crack growth. In real life scenarios, a combination of the two is often observed on the fracture surface.  The presence of each mechanism can attest to the material’s inherent microstructure, the strength of the grain boundaries and the conditions which led to its failure. For example, when exposed to corrosive agents, ceramics often exhibit a transition toward intergranular dominated crack growth, while rapid cracks are more prone to exhibit transgranular fracture.  Unfortunately, the process of quantifying the relative occurrence of each mechanism as observed on the fracture surface is cumbersome and highly user biased. The existence of two different modes and the need of topographic characterization of the fractures is what makes the task at hand challenging. Moreover, when going from one material system to the other, the size of the features on the fracture surface as well as their height fluctuations can vary drastically. Simple classification or object detection algorithms based on deep learning methods are not able to effectively tackle these challenges. The capability of the semantic segmentation algorithms to classify every pixel in the SEM images allows the topographic characterization of the fracture surface, making this approach ideal for a fractographic analysis. Several network architectures were considered and the U-net architecture  (see Methods section) was found to yield slightly better results and hence, the results presented in this work are the predictions of the U-net algorithm on the images of the test dataset. Fig. \ref{sem_example_image} shows one of the SEM images used for the training of the network and the corresponding annotation.

The objective of the training dataset annotation process is to classify the areas of the SEM images that presented the most characteristic features of each micro-fracture mode, while the areas with unclear classification or ambiguous features are labeled as background. This partial annotation was inevitable. The results presented herein can be improved, given a larger, annotated, training set. All of the data used to train the CNN presented here, as well as the source code is freely available on-line with the aim of establishing a large database of fracture surfaces and extending our work to other materials and failure mechanisms.   

During the training process, the algorithm uses the categorical cross-entropy loss in order to determine the deviation between the network predictions and the ground truth data provided by us with the annotated images. The gradient of the loss with respect to the weights is used to perform the updating of the weights during the back-propagation stage of the training. Recording the evolution of the loss allows us to monitor the training process. Similarly, at the end of each epoch the categorical cross-entropy loss of the algorithm's predictions on the validation dataset is computed. The calculation of the prediction accuracy from the loss values is straightforward. It is important to stress that these accuracy values are with respect to the partial annotations that we have manually created from the SEM images. Hence, they do not represent the exact accuracy of the network predictions, but they constitute a very important indication of the efficiency of the algorithm. The training accuracy saturates around a mean value of 72.5\%, while the accuracy on the validation dataset is approximately 71\%. 

After the completion of the training, the trained weights of each layer of the network are exported and saved. Importing these trained weights to a prediction algorithm enables the classification of every pixel in a new SEM image of the fracture surface. The test dataset of the SEM images is used to perform the final evaluation of the algorithm. Following the standard procedures for evaluation of an algorithm in machine learning, the test images have not been used during the training. Although, the network has been trained with images of size $640\times640$ pixels, it is capable of performing predictions in SEM images of any size, as long as the GPU or CPU memory of the computer can handle the image size. 

\begin{figure}
	\centering
    	\subfigure{\includegraphics[width = 40mm]{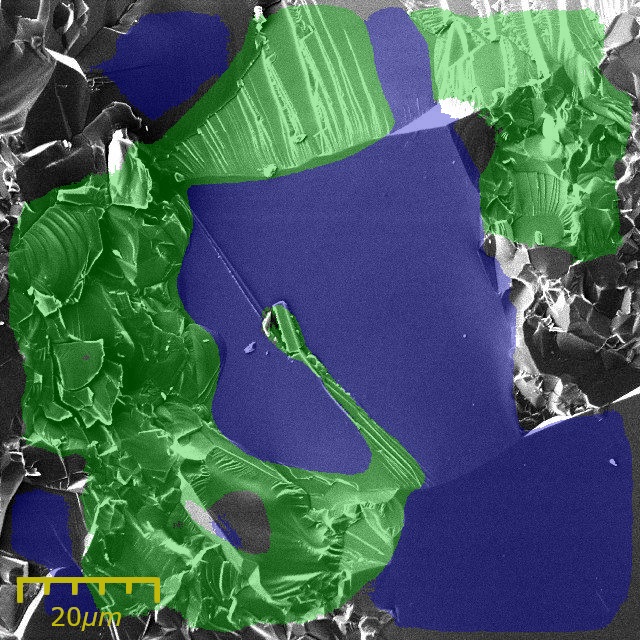} }
		\subfigure{\includegraphics[width = 40mm]{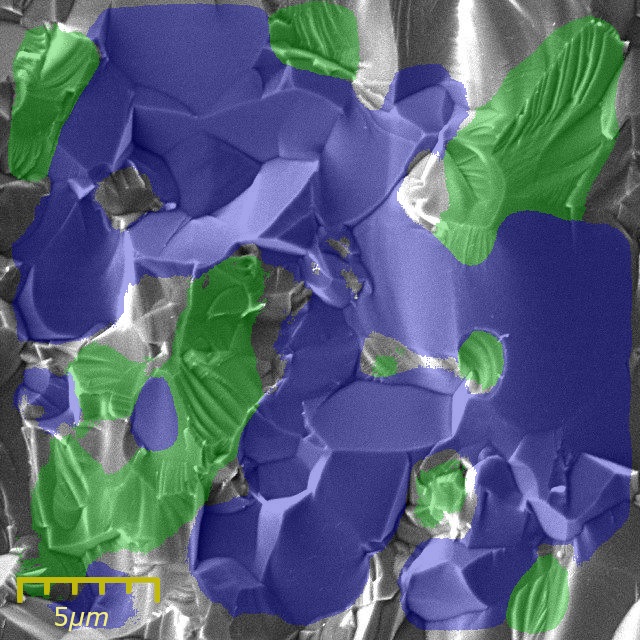}} 
		\subfigure{\includegraphics[width = 40mm]{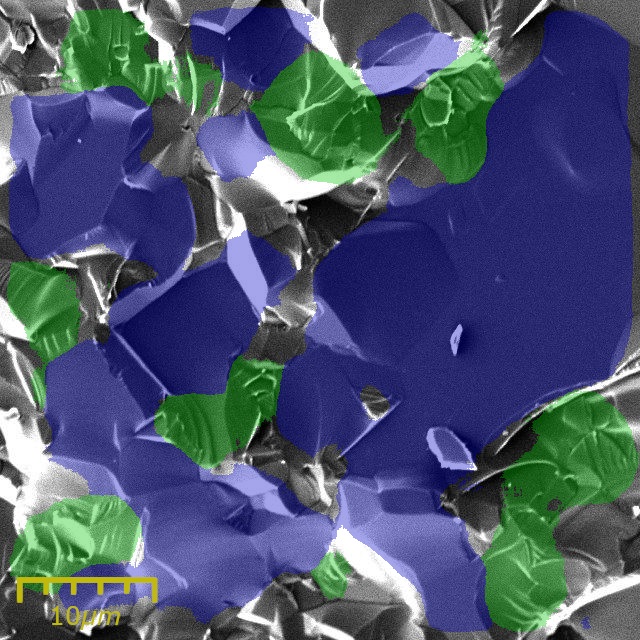}}
		\subfigure{\includegraphics[width = 40mm]{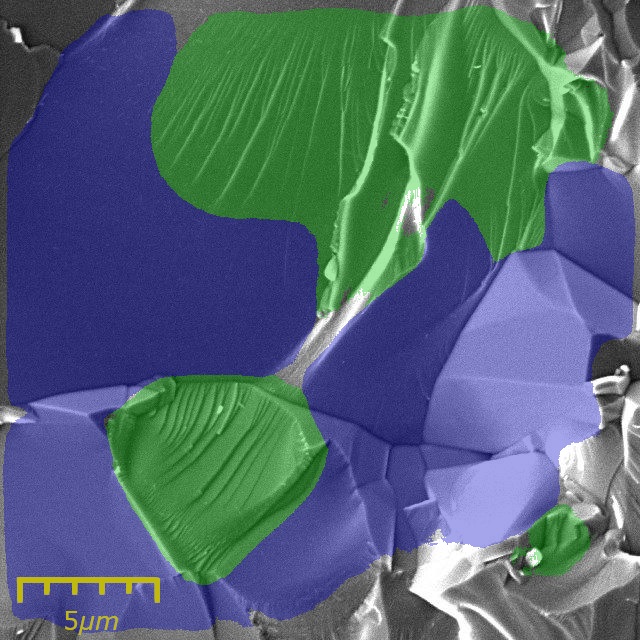}} 
		
		\subfigure{\includegraphics[width = 80mm]{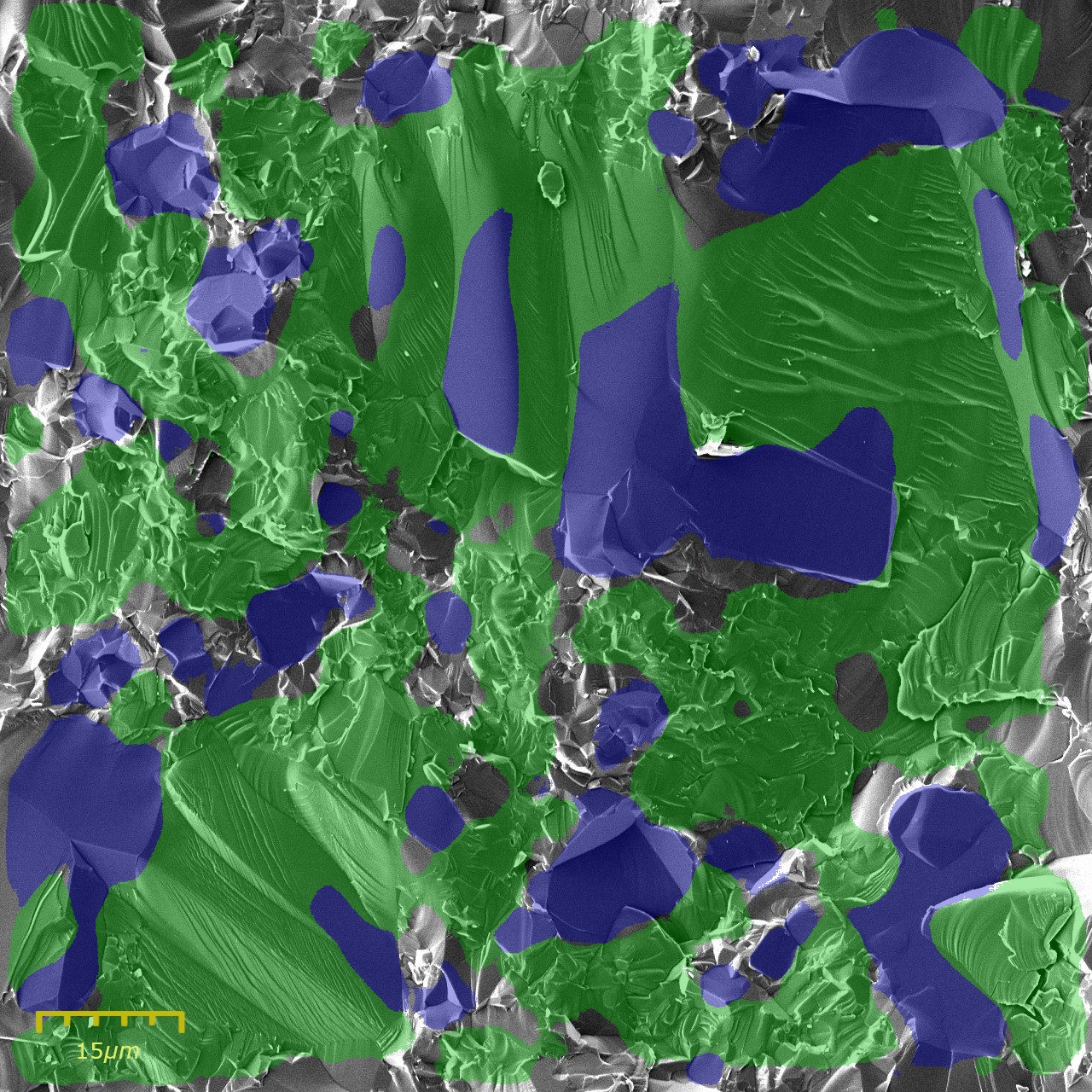}}
			\hspace{1mm}
		\subfigure{\includegraphics[width = 80mm]{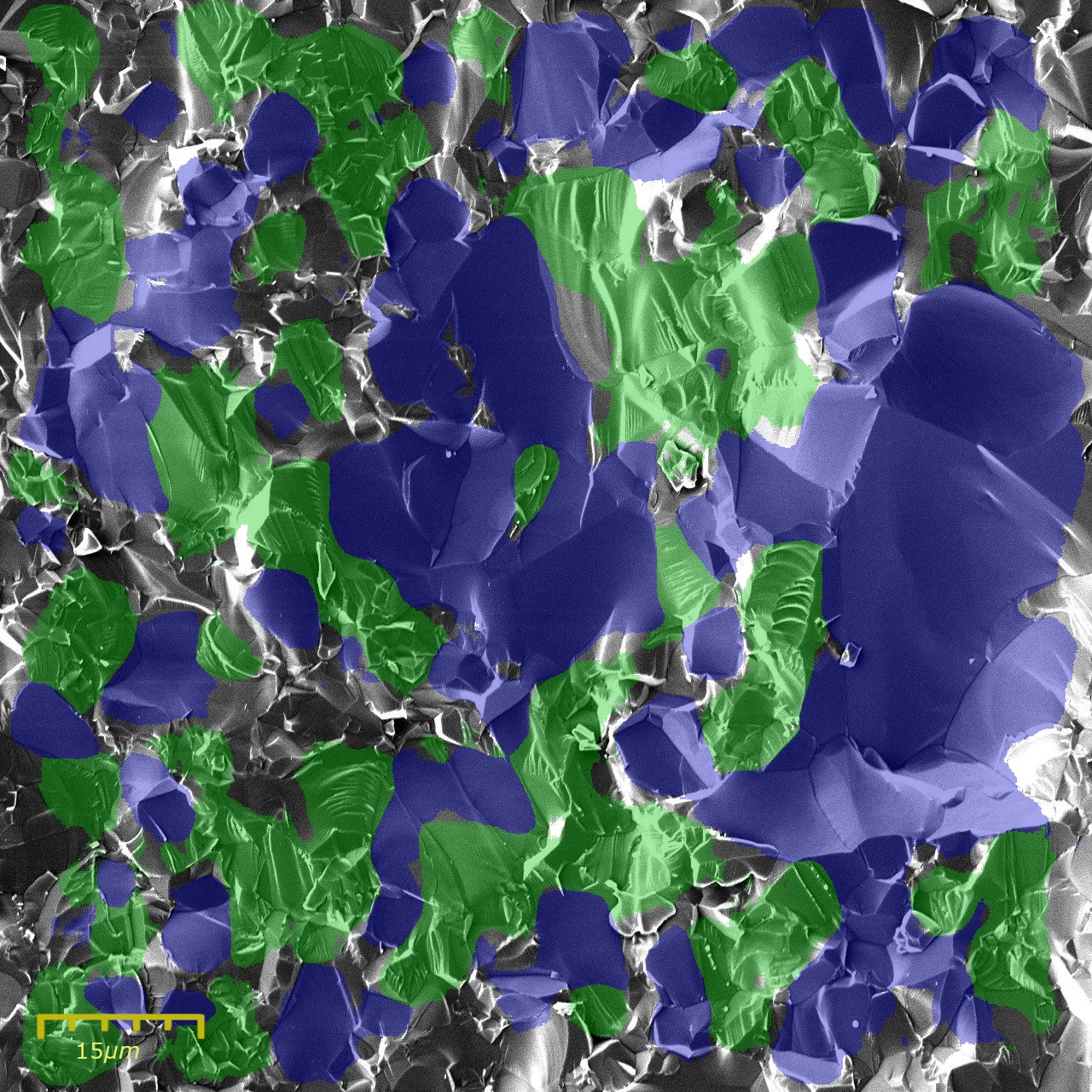}}
			\hspace{1mm}
		\subfigure{\includegraphics[width = 80mm]{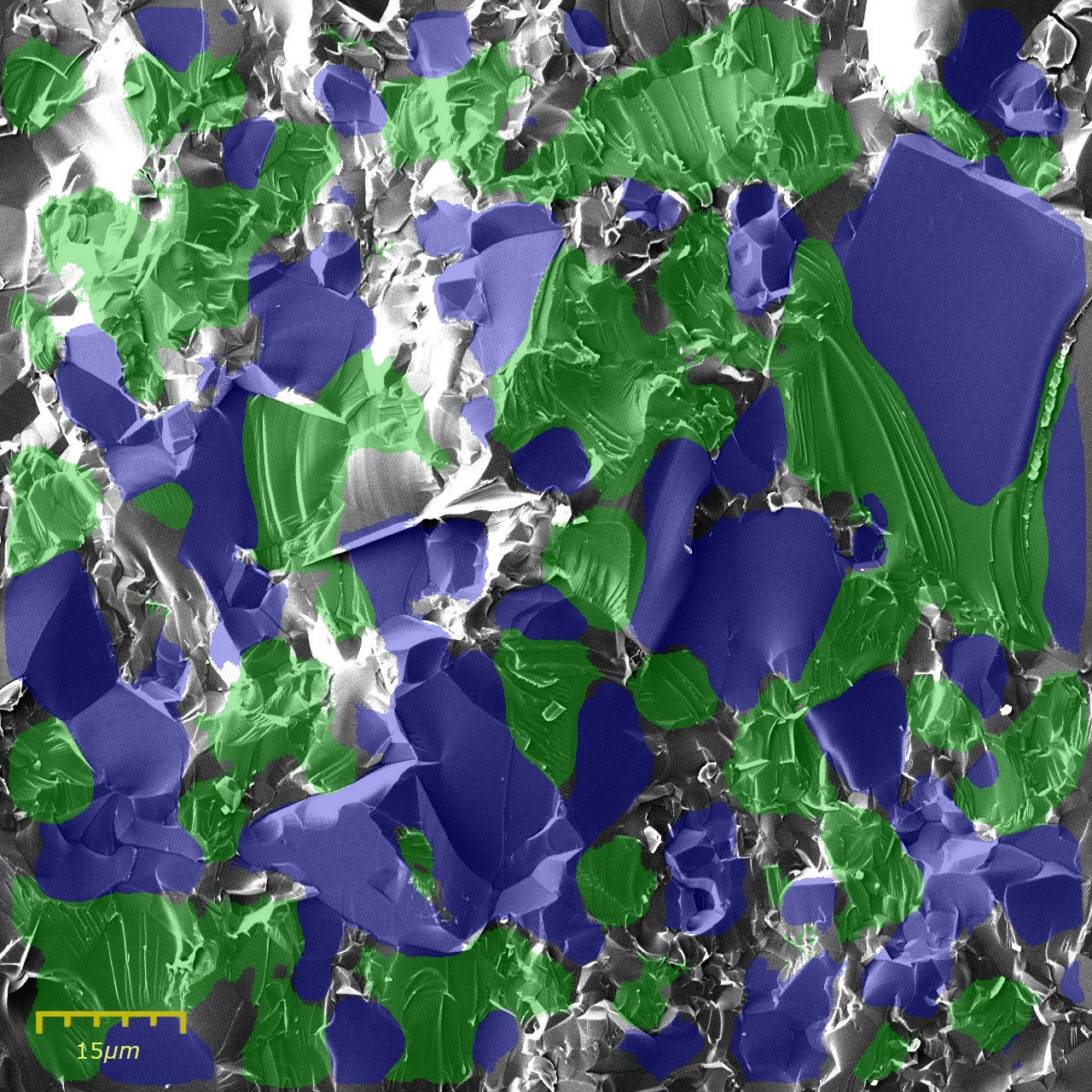}}
			\hspace{1mm}
		\subfigure{\includegraphics[width = 80mm]{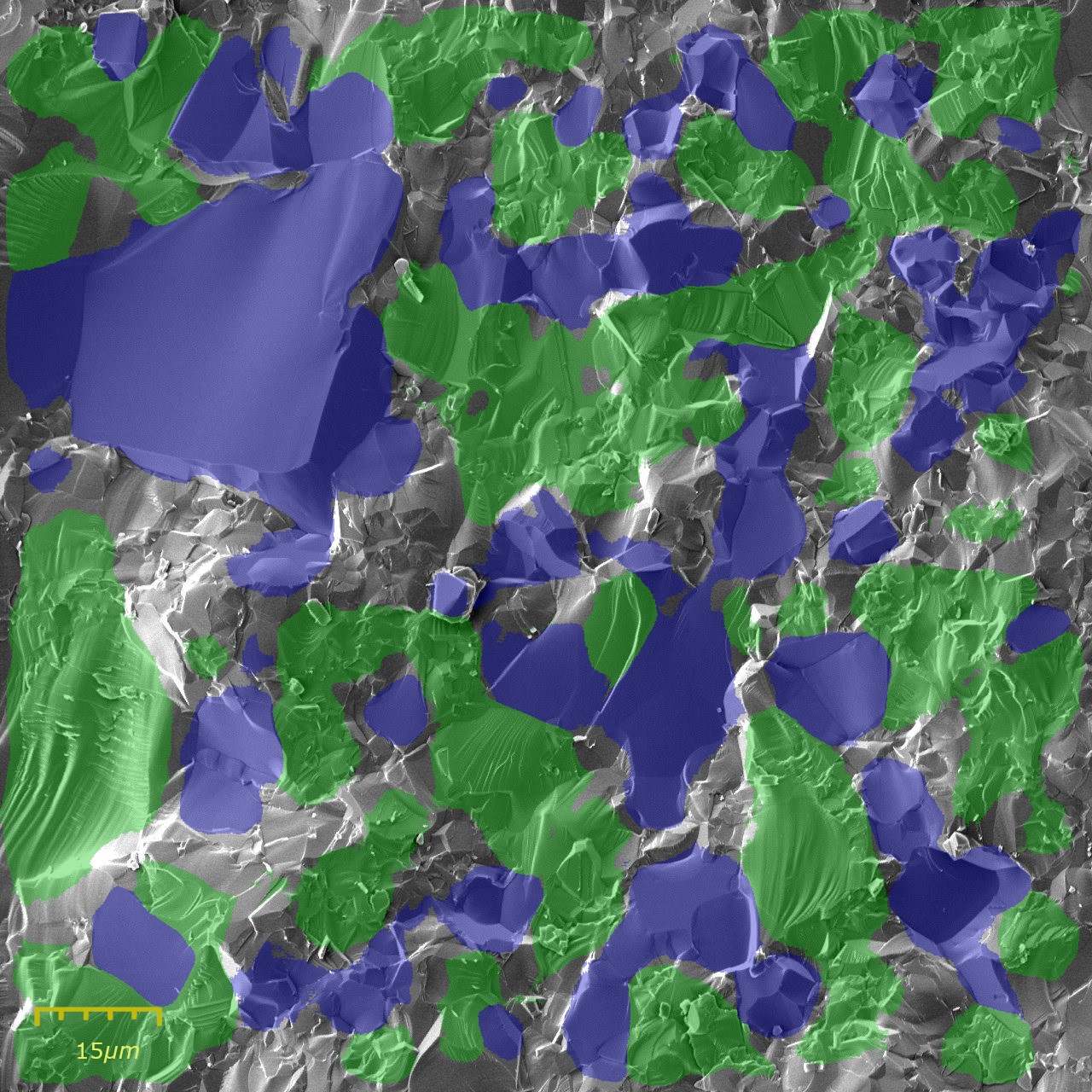}}
	\caption{Algorithm classification results for 8 images selected from the test dataset. Blue colored areas were identified as intergranular while green colored areas correspond to transgranular. Images 1-4 are   $640\times640$ pixels while images 5-8 are with $1280\times1280$ pixels. }
	\label{small_images}
\end{figure}

Figure \ref{small_images} shows the predictions of the algorithm on the small images ($640\times640$ pixels) and on larger images ($1280\times1280$ pixels) of the test dataset. The areas with the blue color are the \textit{intergranular} fracture modes, while the green areas are the \textit{transgranular} modes.  The areas that were left uncolored represent the areas of the \textit{background} class.

From Figure \ref{small_images} it is evident that the semantic segmentation algorithm is able to classify a large percentage of the images' area with high accuracy. The image size does not influence the classification efficiency. In fact, the classifications performed for the larger images exhibit higher accuracy, even though the training is performed on the smaller images. The fact that the classified areas do not cover the entire surface of the image is easily rationalized if we consider that the annotated areas of the images used for the training also do not cover the entire image. The introduction of the \textit{background} class was necessary since manually annotating every pixel of the training and validation dataset is very time-consuming, however it reduces the prediction efficiency. The algorithm tries to learn how to classify the \textit{background} pixels even though these pixels do not follow a certain pattern, which leads to misclassifications. Additionally, there exist regions (e.g. pixels with high brightness) which cannot be classified into one of the two fracture modes either due to the lack of texture or due to their ambiguity even in the eyes of the human user.

\begin{figure}
	\centering
	\subfigure[]{\label{full_annot_image}\includegraphics[width=70mm,height=70mm]{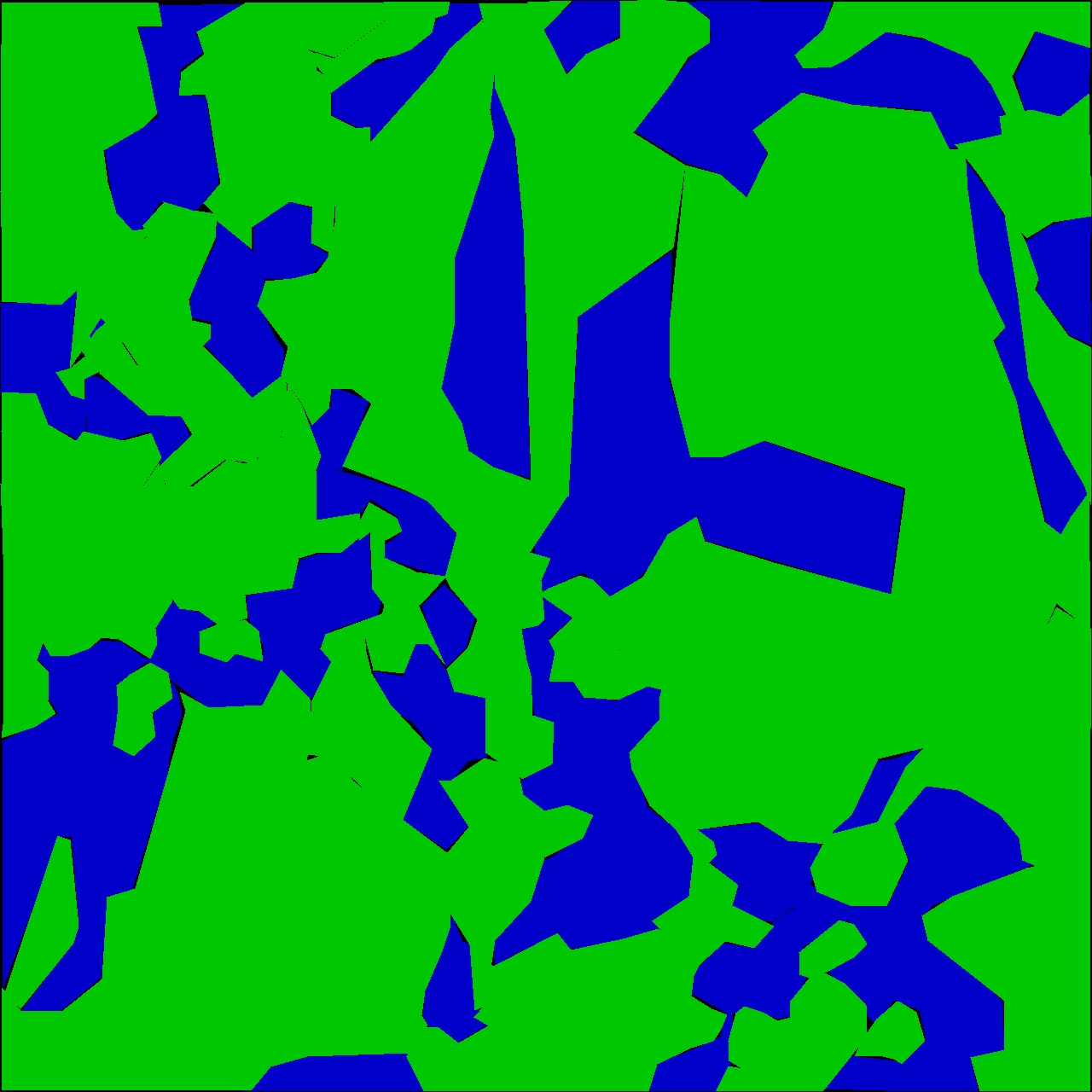}}
	\hspace{1mm}
	\subfigure[]{\label{pred_mask}\includegraphics[width=70mm,height=70mm]{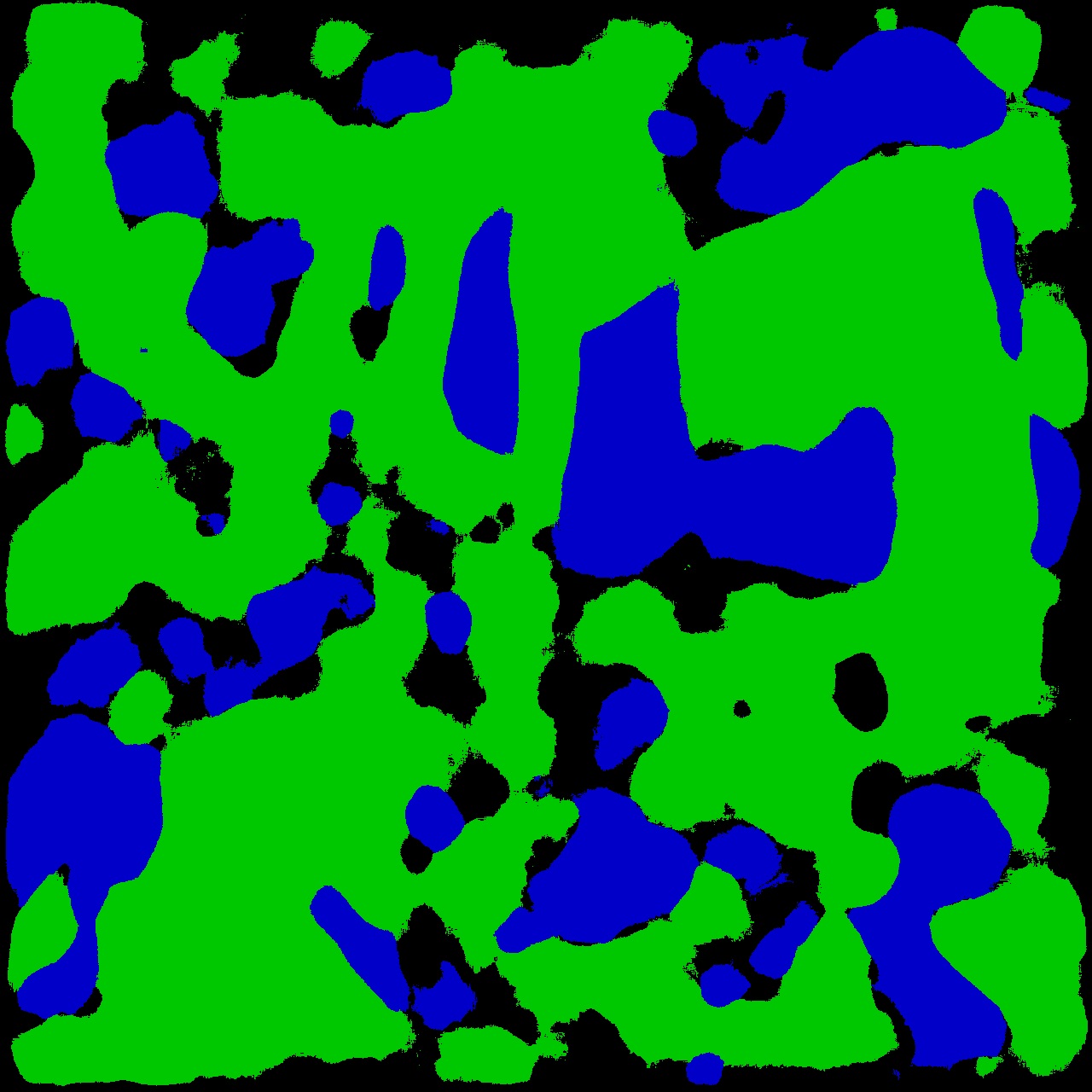}}
	\caption{A fully-annotated image (a) and the corresponding classification mask (b). The IoU was calculated using 8 pairs of fully annotated and classified images similar to the pair shown here.}
	\label{pred_annot_compare}
\end{figure}

In order to evaluate more accurately the efficiency of the algorithm, we decided to perform one additional test. The 8 images of the test dataset presented here have been annotated, but this time we manually labeled every pixel as \textit{intergranular} or \textit{transgranular}, omitting the \textit{background} labeling that was applied during training. An example of this fully-annotated image can be seen in Fig.\ref{full_annot_image}, while in Fig.\ref{pred_mask} we show the algorithm's classification mask of this particular image. The small number of these images permitted such classification. Additionally, the test images have been filtered in order to remove the pixels with high brightness (pixel intensity $>$ 220). The calculation of the Intersection over Union (IoU) between the classified and the labeled image (ground truth) is a more precise measurement of the accuracy of the network's ability to classify pixels in the image as belonging to one of the two fracture modes.     

The IoU is defined as the fraction of the overlap area between the annotations and the predictions of each fracture mode and the union area between them. The mean value of the IoU for the \textit{intergranular} mode for the eight prediction images is 77.3\%, while for the \textit{transgranular} mode is 63.3\%. The total mean accuracy is 71.2\%, which is very close to the validation accuracy that was computed during training. 

The evaluation of the capability of the algorithm in predicting correctly the fracture mode of each pixel is obscured by the introduction of the \textit{background} labeling. The existence of a label, which does not express any fracture mode, does not allow the correct evaluation of the algorithm's accuracy. This is not an unusual issue in semantic segmentation algorithms\cite{Thoma2016}. To address this issue a \textit{void} class is introduced and assigned for each pixel that has not been annotated in the training dataset\cite{Gould2008,Thoma2016}. Similar approach was followed by the Microsoft research group \cite{Shotton2006} where the \textit{void} labeling was used for pixels located in the boundaries of different classes or in general imposed difficulties in labeling, making the annotation process very slow. In both cases these pixels have been excluded from the ground truth during the evaluation of the accuracy in the test dataset predictions. 

In our case, these pixels are the \textit{background} pixels, and when excluded from the ground truth images, the IoU values increase significantly. The IoU of the \textit{intergranular} and the \textit{transgranular} mode for the same test images become 93\% and 87.4\%, respectively. The total mean IoU value of the algorithm raises to 91.1\%.  

Finally, the F-measure of this binary classification is computed using the following formula \cite{Thoma2016}:

\begin{equation}
F_{\beta} = \left( 1 + \beta^2\right) \frac{\textrm{tp}}{\left( 1 + \beta^2\right)\textrm{tp} + \beta^2\textrm{fn} + \textrm{fp}}
\end{equation}

where we consider as positive the \textit{intergranural} pixels and negative the \textit{transgranural} pixels. The tp, fp and fn are the true positive, false positive and false negative pixels and $\beta$ is set to 1. 

The resulting F-measure is 90.7\%, and it is in agreement with the computed mean IoU value. 
\subsection{Transferability} \hfill

Once the functionality and accuracy of the algorithm in the test dataset of the fracture SEM images of the $MgAl_2O_4$ samples was evaluated, the next step is the investigation of the transferability of the algorithm and trained network to a different ceramic material, for which no additional training was performed. To this end, we have obtained $Al_2O_3$ broken samples and, subsequently, SEM images of the fracture surface were acquired. $Al_2O_3$ is a brittle material, but the fracture surface has different characteristics and morphology than the previously studied $MgAl_2O_4$ fracture surface. Since the classification on the new images is performed using the previously trained weights, we were able  to asses the ability of the trained network to accurately classify \textit{intergranular} and \textit{transgranular} regions in materials for which it was not trained. 

\begin{figure}
	\centering
	\begin{tabular}{c c c}
		\includegraphics[width = 80mm]{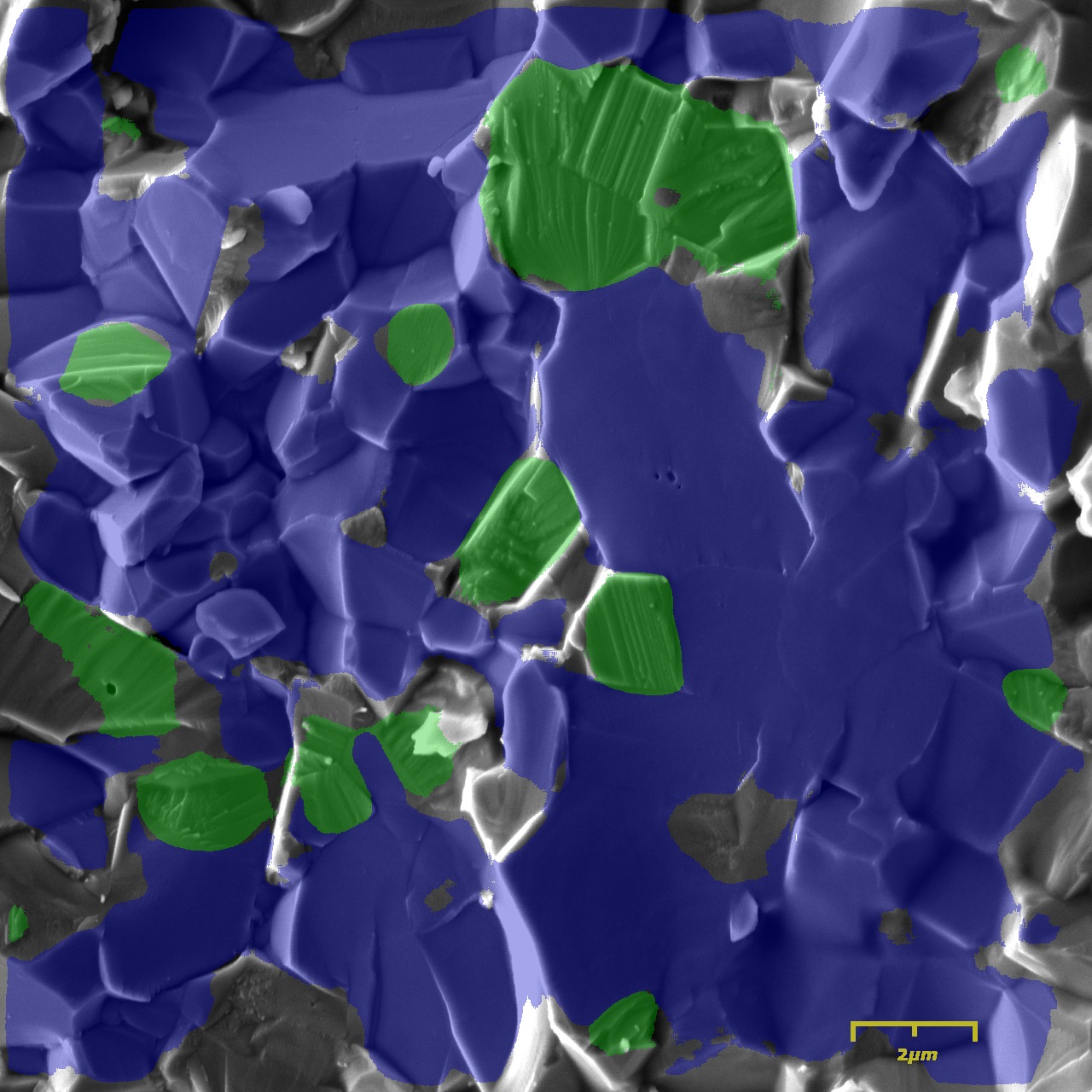} &
		\includegraphics[width = 80mm]{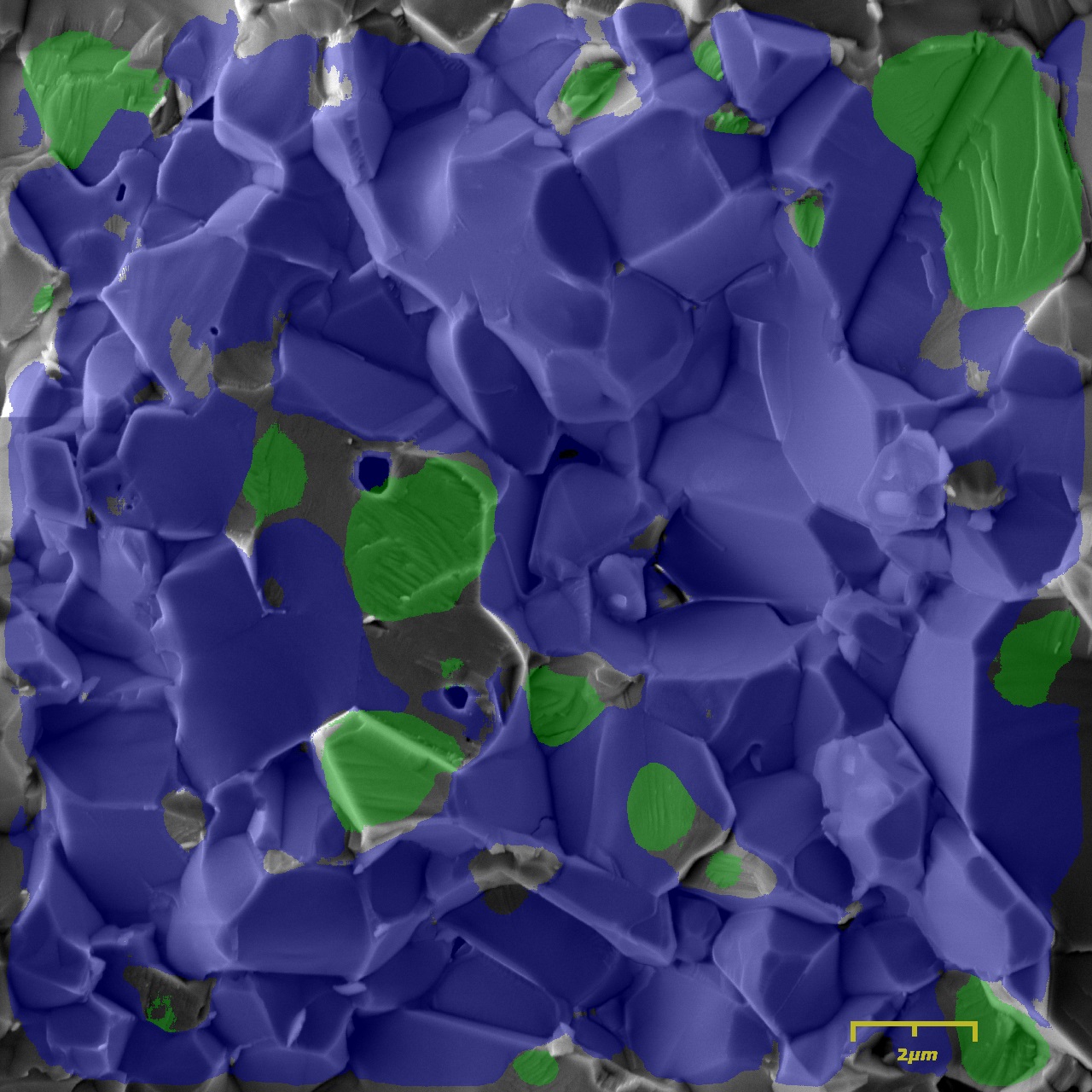} \\
		\includegraphics[width = 80mm]{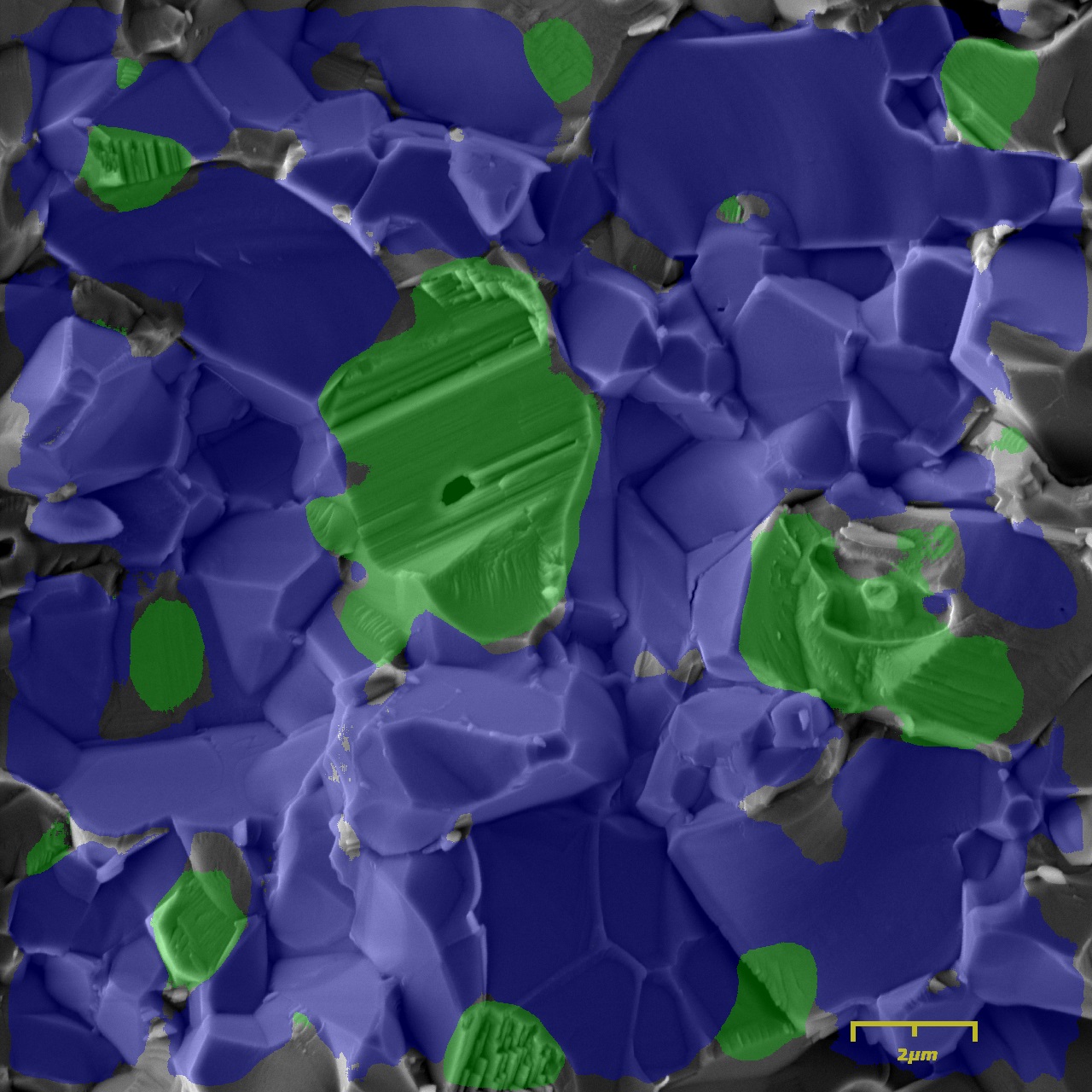} &
		\includegraphics[width = 80mm]{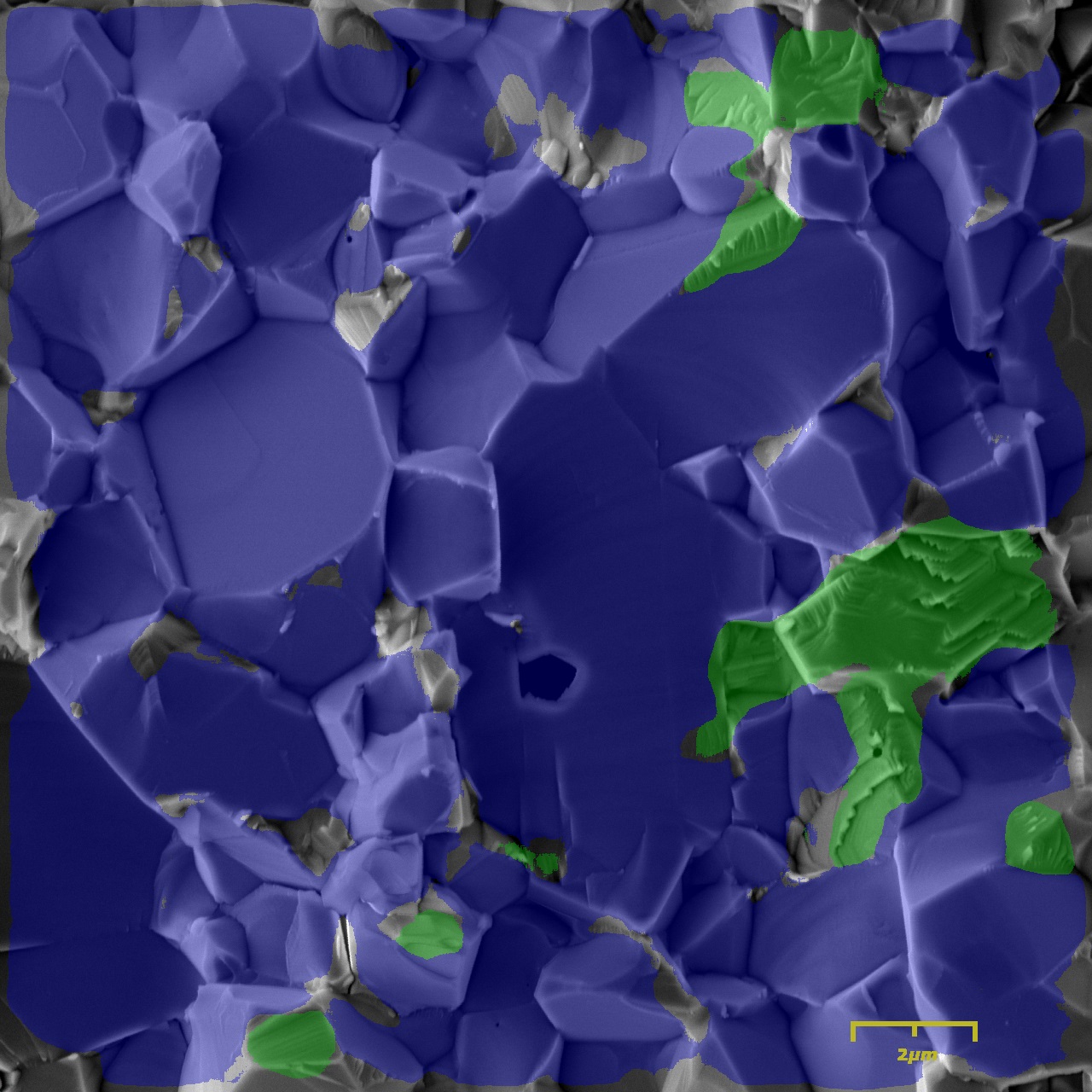} \\
	\end{tabular}
	\caption{Classification of the SEM images of the $Al_2O_3$ fracture surfaces, with size of $1280\times1280$ pixels. It is important to note that these results are obtained without additional training of the network.}
	\label{large_images_Al2O3}
\end{figure}

The test dataset for the new samples consisted of 6 SEM images cropped to a size of $1280\times1280$ pixels. The performance evaluation on the $Al_2O_3$ was conducted over the same overall number of pixels as in the original test dataset ($MgAl_2O_4$). In Fig.\ref{large_images_Al2O3}, we present the classification of four SEM images of the $Al_2O_3$ fracture surfaces.

To evaluate the accuracy of the algorithm on these new fracture images we used the same tools as before.  Initially, the SEM images were filtered by removing the high brightness areas and the mean value for the total IoU accuracy is computed to be 78\%. Subsequently, following the evaluation methodology applied for the $MgAl_2O_4$ test dataset, we removed the \textit{void} (or \textit{background}) annotated pixel areas and we computed again the mean value of the total IoU and the F-measure, and the results are 94\% and 82.4\%, respectively.    

These results show that the performance of the algorithm in a new material, without any additional training, remained highly accurate. Nevertheless, we acknowledge that a more extensive testing of the algorithm in different ceramic material samples will provide a clearer picture. Furthermore, it is important to add that it will be straightforward to identify which features are completely new in any new dataset and the previous training can be easily optimized by only adding annotations of features that  differentiate the new dataset from the previous training dataset. With this training methodology, the optimization of the algorithm will not require a new extensive annotated dataset and it will not be as time-consuming as the initial training.   

\section*{Summary}
\label{conclusions_label}
In this work, we have utilized a pre-trained convolutional neural network designed for semantic segmentation purposes (U-net architecture) to study the fracture surface of $MgAl_2O_4$ samples. By fine tuning the weights of the various layers of the network we were able to achieve a total mean IoU value of $91.1\%$ on the SEM test dataset. 
To test the flexibility and robustness of this approach, additional broken samples of Alumina ($Al_2O_3$) were imaged and subjected to the same analysis without further training of the model for this specific materials. Despite the large differences in the microstructural feature sizes, the performance of the algorithm remained high, exhibiting a total mean IoU value of $94\%$, possibly due to the large dominance of the intergranular fracture which is better classified in our training set. 

The approach presented here, provides a robust and user independent tool for quantitative analysis of ceramic’s fracture surfaces, which can be utilized to study large areas on the fracture surface with ease.  The proposed method can be easily integrated into the engineering failure analysis process, independently of the expertise level of the failure analysis engineer.

While our preliminary investigation suggests that a good transition between different material systems is possible, even in the case of major changes in the morphological characteristics of the fracture surface it is safe to hypothesize that only a small additional training phase on the newly observed feature is required to further fine tune the model. Furthermore, the extension of the model to accommodate other classes of features, (e.g. dimples) seems to be straightforward by introducing and training the algorithm to recognize the new class. A study as to the feasibility and robustness of this hypothesis is the objective of our future work.

\begin{methods}

\subsection{Network Architecture.}

In the work presented here, the encoder part of the network is constructed according to the architecture of the VGG16 network without including the last two fully connected layers and the Softmax classifier. The building unit of the VGG network is a 3x3 convolutional layer that is repeated 2 or 3 times, depending on the layer position in the network, followed by a ReLu activation \cite{Nair2010} layer and a 2x2 stride max pooling layer. The total amount of layers in this network is 16, hence the name VGG16. Depending on the decoder architecture but also on the connections between the encoder and decoder, different architectures are proposed in the scientific literature. We have implemented three different network architectures, which are considered the most efficient for the task of semantic segmentation \cite{Long2015, Badrinarayanan2017, Ronneberger2015}, and after comparing their efficiency we chose the U-net that performed better for the task at hand. Additionally, the ability of this architecture to achieve high accuracy training with small datasets was a very compelling feature.

The code is developed in Python on top of an existing implementation  (https://github.com/ divamgupta/image-segmentation-keras),  and the building and training of the model is performed with Keras \cite{chollet2015keras} Application Programming Interface(API), using Tensorflow \cite{Abadi2016,tensorflow2015-whitepaper} as a backend.

The SEM input images used for the training of the network have a size of $640\times640$ pixels, and after applying the consecutive convolution and max pooling layers of the encoder the size of the resulting feature maps is reduced to $20\times20$ pixels. The role of the decoder is to retrieve the accurate localization of these features and convert the feature maps into pixel-wise predictions on the input images. The main tool of this upsampling process is the transpose convolution (or deconvolution) layer. These deconvolution layers perform the reverse operation of the typical convolution layers and their weights are trainable parameters of the network, which allows the network to adjust them during the training and learn how to accurately localize the features in any new input image. 

The encoder and decoder structure of the U-net architecture is identical, with the only difference being the replacement of the max pooling operations in the encoder stage with a 2x2 transpose convolutions in the decoder stage. The symmetrical architecture of the network allows the concatenation of the output of each encoder layer with the output of the corresponding decoder layer, before being fed to the next decoder layer. With these modifications, the network can utilize larger number of feature channels and subsequently propagate more information through the layers and producing higher resolution feature maps. This interconnection between the encoder and decoder also allows the network to be effectively trained with a smaller dataset.        

\subsection{SEM imaging and Training.}
The training dataset is constructed using SEM images of the fracture surface of Magnesium Aluminate Spinel ($MgAl_2O_4$) samples. The fracture experiments were conducted by D. Blumer\cite{blumer}. The fracture surfaces were prepared and scanned using an high resolution FEG-SEM( TESCAN MIRA 3), following the general guidelines for fractography of ceramics \cite{quinn2016}. All images were taken at a resolution of $2560\times2560$ pixels and subsequently cropped to desired dimensions. For the purpose of constructing a training dataset, the images were cropped to $640\times640$ pixels size. The final dataset is composed of 605 training images, 105 validation images and 30 test images (used for the final assessment of the prediction accuracy of each algorithm). The training and validation images are annotated using the open source online VGG Image Annotator \cite{dutta2019vgg,dutta2016via} in three different labels: \textit{intergranular} , \textit{transgranular} and \textit{background}. The annotation was a manual time-consuming task and the inclusion of the third classification label (the \textit{"background"} label) was a necessity in order to simplify this rigorous process. Moreover, ambiguities as to the nature of several features were eliminated by annotating them using the \textit{"background"} class.  In the resulted dataset, each annotated image contains on average 40 classified areas of \textit{intergranular} and \textit{transgranular} fracture modes, while the rest of the pixels are classified as \textit{"background"}. 

The weights of each algorithm are trained for 40 epochs with Adam Stochastic gradient-based optimization method \cite{Kingma2014}, with initial learning rate of $10^{-6}$, $\beta_1 = 0.9$ and $\beta_2 = 0.999$. Each epoch consisted of 200 iterations followed by 100 validation iterations. The training was performed on a personal computer, equipped with an NVIDIA\textsuperscript{\textregistered} GeForce\textsuperscript{\textregistered} RTX 2080 Ti  Graphics Processing Unit(GPU). Taking advantage of the GPU support of the Tensorflow operations, the training duration is significantly reduced. For a relatively small batch size (batch size = 4), since the memory size of our GPU is limited, the training time for the different networks was approximately 150 sec per epoch. 

\subsection{Materials.}
 Fractured specimens made of transperent ($MgAl_2O_4$) Spinel specimens were acquired from the dynamic fracture laboratory at the Technion\cite{blumer}. The ($MgAl_2O_4$) specimens consisted of two populations in terms of microstructure. The first exhibited a bi-modal grain size distribution with sub-micron grains and grains at the order of several microns. The second population has gone through a thermomechanical treatment to facilitate abnormal grain growth, resulting in the appearance of several grains as large as 140 microns\cite{blumer}. 
 The ($Al_2O_3$) samples used for this study are sintered specimens with 2\% porosity. The average grain size was $\sim$ 4 microns.
\subsection{Data availability}
The training and validation dataset used for the training of the network and their annotations are published in Materials Data Facility(MDF) with DOI: \textit{https://doi.org/10.18126/vu60-4htj}. The source code is available at \textit{https://github.com/SteliosTsop/QF-image-segmentation-keras}.  

\end{methods}

\section*{Acknowledgments}
	The financial support provided by the Pazy foundation young researchers award Grant $\#$ 1176 (SO) and the European Union's Horizon2020 Programme (Excellent Science, Marie-Sklodowska-Curie Actions) under REA grant agreement $675602$ (Project OUTCOME, TS, Rxx, SO), is gratefully acknowledged. The authors would also like to thank Aerosertec(\textit{www.aerosertec.com}) for providing the facilities and computational power for the training of the network.Finally, we would like to thank Dr. D.Rittel and D.Blumer for providing the Magnesium Aluminate Spinel ($MgAl_2O_4$) fractured samples.

\section*{Author contributions}
S.O conceived the research. S.O and S.T performed the fractography and wrote the manuscript. R.H.M. was an industrial advisor to the project. Coding, annotations and fine tuning of the CNN were done by S.T.
\begin{addendum}
	\item[Competing Interests] The authors declare that they have no
	competing financial interests.
\end{addendum}


\section*{References}
\bibliography{quantfrac}

\begin{thebibliography}{10}
\expandafter\ifx\csname url\endcsname\relax
  \def\url#1{\texttt{#1}}\fi
\expandafter\ifx\csname urlprefix\endcsname\relax\def\urlprefix{URL }\fi
\providecommand{\bibinfo}[2]{#2}
\providecommand{\eprint}[2][]{\url{#2}}

\bibitem{ASM}
\bibinfo{author}{Mills, K.}, \bibinfo{author}{Davis, J.~R.},
  \bibinfo{author}{Destefani, J.} \& \bibinfo{author}{Dieterich, D.}
\newblock \emph{\bibinfo{title}{ASM Handbook, Volume 12-Fractography}}
  (\bibinfo{publisher}{Asm International Materials Park, OH},
  \bibinfo{year}{1987}).

\bibitem{tipper}
\bibinfo{author}{Tipper, C.}
\newblock \bibinfo{title}{The fracture of metals}.
\newblock \emph{\bibinfo{journal}{Metallurgia}} \textbf{\bibinfo{volume}{39}},
  \bibinfo{pages}{133--137} (\bibinfo{year}{1949}).

\bibitem{pineau1}
\bibinfo{author}{Pineau, A.}, \bibinfo{author}{Benzerga, A.~A.} \&
  \bibinfo{author}{Pardoen, T.}
\newblock \bibinfo{title}{Failure of metals i: Brittle and ductile fracture}.
\newblock \emph{\bibinfo{journal}{Acta Materialia}}
  \textbf{\bibinfo{volume}{107}}, \bibinfo{pages}{424--483}
  (\bibinfo{year}{2016}).

\bibitem{pineau2}
\bibinfo{author}{Pineau, A.}, \bibinfo{author}{McDowell, D.~L.},
  \bibinfo{author}{Busso, E.~P.} \& \bibinfo{author}{Antolovich, S.~D.}
\newblock \bibinfo{title}{Failure of metals ii: Fatigue}.
\newblock \emph{\bibinfo{journal}{Acta Materialia}}
  \textbf{\bibinfo{volume}{107}}, \bibinfo{pages}{484--507}
  (\bibinfo{year}{2016}).

\bibitem{marbel}
\bibinfo{author}{Zhang, Q.~B.} \& \bibinfo{author}{Zhao, J.}
\newblock \bibinfo{title}{Effect of loading rate on fracture toughness and
  failure micromechanisms in marble}.
\newblock \emph{\bibinfo{journal}{Engineering Fracture Mechanics}}
  \textbf{\bibinfo{volume}{102}}, \bibinfo{pages}{288--309}
  (\bibinfo{year}{2013}).

\bibitem{hu2012mechanisms}
\bibinfo{author}{Hu, G.}, \bibinfo{author}{Chen, C.}, \bibinfo{author}{Ramesh,
  K.} \& \bibinfo{author}{McCauley, J.}
\newblock \bibinfo{title}{Mechanisms of dynamic deformation and dynamic failure
  in aluminum nitride}.
\newblock \emph{\bibinfo{journal}{Acta materialia}}
  \textbf{\bibinfo{volume}{60}}, \bibinfo{pages}{3480--3490}
  (\bibinfo{year}{2012}).

\bibitem{keren}
\bibinfo{author}{Shemtov-Yona, K.}, \bibinfo{author}{{\"O}zcan, M.} \&
  \bibinfo{author}{Rittel, D.}
\newblock \bibinfo{title}{Fractographic characterization of fatigued zirconia
  dental implants tested in room air and saline solution}.
\newblock \emph{\bibinfo{journal}{Engineering Failure Analysis}}
  \textbf{\bibinfo{volume}{96}}, \bibinfo{pages}{298--310}
  (\bibinfo{year}{2019}).

\bibitem{sigma}
\bibinfo{author}{Biezma, M.~V.}, \bibinfo{author}{Berlanga, C.} \&
  \bibinfo{author}{Argandona, G.}
\newblock \bibinfo{title}{Relationship between microstructure and fracture
  types in a uns s32205 duplex stainless steel}.
\newblock \emph{\bibinfo{journal}{Materials Research}}
  \textbf{\bibinfo{volume}{16}}, \bibinfo{pages}{965--969}
  (\bibinfo{year}{2013}).

\bibitem{molinari}
\bibinfo{author}{Kraft, R.} \& \bibinfo{author}{Molinari, J.}
\newblock \bibinfo{title}{A statistical investigation of the effects of grain
  boundary properties on transgranular fracture}.
\newblock \emph{\bibinfo{journal}{Acta Materialia}}
  \textbf{\bibinfo{volume}{56}}, \bibinfo{pages}{4739--4749}
  (\bibinfo{year}{2008}).

\bibitem{soijf}
\bibinfo{author}{Osovski, S.}, \bibinfo{author}{Needleman, A.} \&
  \bibinfo{author}{Srivastava, A.}
\newblock \bibinfo{title}{Intergranular fracture prediction and microstructure
  design}.
\newblock \emph{\bibinfo{journal}{International Journal of Fracture}}
  \textbf{\bibinfo{volume}{216}}, \bibinfo{pages}{135--148}
  (\bibinfo{year}{2019}).

\bibitem{Hu2017}
\bibinfo{author}{Hu, W.}, \bibinfo{author}{Wiliem, A.},
  \bibinfo{author}{Lovell, B.}, \bibinfo{author}{Barter, S.} \&
  \bibinfo{author}{Liu, L.}
\newblock \bibinfo{title}{{Automation of Quantitative Fractography for
  Determination of Fatigue Crack Growth Rates with Marker Loads}}.
\newblock In \emph{\bibinfo{booktitle}{29th ICAF Symposium – Nagoya}},
  \bibinfo{number}{June} (\bibinfo{year}{2017}).

\bibitem{Kosarevych2013}
\bibinfo{author}{Kosarevych, R.~Y.}, \bibinfo{author}{Student, O.~Z.},
  \bibinfo{author}{Svirs'Ka, L.~M.}, \bibinfo{author}{Rusyn, B.~P.} \&
  \bibinfo{author}{Nykyforchyn, H.~M.}
\newblock \bibinfo{title}{{Computer analysis of characteristic elements of
  fractographic images}}.
\newblock \emph{\bibinfo{journal}{Materials Science}}
  \textbf{\bibinfo{volume}{48}}, \bibinfo{pages}{474--481}
  (\bibinfo{year}{2013}).

\bibitem{Kenjiro1993}
\bibinfo{author}{Kenjiro, K.}, \bibinfo{author}{Minoshima, K.} \&
  \bibinfo{author}{Shoich, I.}
\newblock \bibinfo{title}{{Recognition of Different Fracture Surface
  Morphologies using Computer Image Processing Technique}}.
\newblock \emph{\bibinfo{journal}{JSME international journal. Ser. A, Mechanics
  and material engineering}} \textbf{\bibinfo{volume}{36}},
  \bibinfo{pages}{220--227} (\bibinfo{year}{1993}).

\bibitem{Dutta2014}
\bibinfo{title}{{Characterization of micrographs and fractographs of
  Cu-strengthened HSLA steel using image texture analysis}}.
\newblock \emph{\bibinfo{journal}{Measurement: Journal of the International
  Measurement Confederation}} \textbf{\bibinfo{volume}{47}},
  \bibinfo{pages}{130--144} (\bibinfo{year}{2014}).
\newblock \urlprefix\url{http://dx.doi.org/10.1016/j.measurement.2013.08.030}.

\bibitem{yang1991sem}
\bibinfo{author}{Yang, W.-J.}, \bibinfo{author}{Yu, C.-T.} \&
  \bibinfo{author}{Kobayashi, A.~S.}
\newblock \bibinfo{title}{Sem quantification of transgranular vs intergranular
  fracture}.
\newblock \emph{\bibinfo{journal}{Journal of the American Ceramic Society}}
  \textbf{\bibinfo{volume}{74}}, \bibinfo{pages}{290--295}
  (\bibinfo{year}{1991}).

\bibitem{Chowdhury2016}
\bibinfo{author}{Chowdhury, A.}, \bibinfo{author}{Kautz, E.},
  \bibinfo{author}{Yener, B.} \& \bibinfo{author}{Lewis, D.}
\newblock \bibinfo{title}{{Image driven machine learning methods for
  microstructure recognition}}.
\newblock \emph{\bibinfo{journal}{Computational Materials Science}}
  \textbf{\bibinfo{volume}{123}}, \bibinfo{pages}{176--187}
  (\bibinfo{year}{2016}).
\newblock \urlprefix\url{http://dx.doi.org/10.1016/j.commatsci.2016.05.034}.

\bibitem{Zhang2019}
\bibinfo{author}{Zhang, Y.} \& \bibinfo{author}{Ngan, A.~H.}
\newblock \bibinfo{title}{{Extracting dislocation microstructures by deep
  learning}}.
\newblock \emph{\bibinfo{journal}{International Journal of Plasticity}}
  \textbf{\bibinfo{volume}{115}}, \bibinfo{pages}{18--28}
  (\bibinfo{year}{2019}).
\newblock \urlprefix\url{https://doi.org/10.1016/j.ijplas.2018.11.008}.

\bibitem{Li2018}
\bibinfo{author}{Li, W.}, \bibinfo{author}{Field, K.~G.} \&
  \bibinfo{author}{Morgan, D.}
\newblock \bibinfo{title}{{Automated defect analysis in electron microscopic
  images}}.
\newblock \emph{\bibinfo{journal}{npj Computational Materials}}
  \textbf{\bibinfo{volume}{4}}, \bibinfo{pages}{1--9} (\bibinfo{year}{2018}).
\newblock \urlprefix\url{http://dx.doi.org/10.1038/s41524-018-0093-8}.

\bibitem{Gola2018}
\bibinfo{author}{Gola, J.} \emph{et~al.}
\newblock \bibinfo{title}{{Advanced microstructure classification by data
  mining methods}}.
\newblock \emph{\bibinfo{journal}{Computational Materials Science}}
  \textbf{\bibinfo{volume}{148}}, \bibinfo{pages}{324--335}
  (\bibinfo{year}{2018}).
\newblock \urlprefix\url{https://doi.org/10.1016/j.commatsci.2018.03.004}.

\bibitem{Bastidas-Rodriguez2016}
\bibinfo{author}{Bastidas-Rodriguez, M.~X.}, \bibinfo{author}{Prieto-Ortiz,
  F.~A.} \& \bibinfo{author}{Espejo, E.}
\newblock \bibinfo{title}{{Fractographic classification in metallic materials
  by using computer vision}}.
\newblock \emph{\bibinfo{journal}{Engineering Failure Analysis}}
  \textbf{\bibinfo{volume}{59}}, \bibinfo{pages}{237--252}
  (\bibinfo{year}{2016}).
\newblock \urlprefix\url{http://dx.doi.org/10.1016/j.engfailanal.2015.10.008}.

\bibitem{Konovalenko2018}
\bibinfo{author}{Konovalenko, I.}, \bibinfo{author}{Maruschak, P.},
  \bibinfo{author}{Prentkovskis, O.} \& \bibinfo{author}{Junevi{\v{c}}ius, R.}
\newblock \bibinfo{title}{{Investigation of the Rupture Surface of the Titanium
  Alloy Using Convolutional Neural Networks}}.
\newblock \emph{\bibinfo{journal}{Materials}} \textbf{\bibinfo{volume}{11}},
  \bibinfo{pages}{2467} (\bibinfo{year}{2018}).

\bibitem{chen2014semantic}
\bibinfo{author}{Chen, L.-C.}, \bibinfo{author}{Papandreou, G.},
  \bibinfo{author}{Kokkinos, I.}, \bibinfo{author}{Murphy, K.} \&
  \bibinfo{author}{Yuille, A.~L.}
\newblock \bibinfo{title}{Semantic image segmentation with deep convolutional
  nets and fully connected crfs}.
\newblock \emph{\bibinfo{journal}{arXiv preprint arXiv:1412.7062}}
  (\bibinfo{year}{2014}).

\bibitem{Simonyan2014}
\bibinfo{author}{Simonyan, K.} \& \bibinfo{author}{Zisserman, A.}
\newblock \bibinfo{title}{{Very Deep Convolutional Networks for Large-Scale
  Image Recognition}}.
\newblock \emph{\bibinfo{journal}{CoRR}} \bibinfo{pages}{1--14}
  (\bibinfo{year}{2014}).
\newblock \urlprefix\url{http://arxiv.org/abs/1409.1556}.
\newblock \eprint{1409.1556}.

\bibitem{He2016}
\bibinfo{author}{He, K.}, \bibinfo{author}{Zhang, X.}, \bibinfo{author}{Ren,
  S.} \& \bibinfo{author}{Sun, J.}
\newblock \bibinfo{title}{{Deep residual learning for image recognition}}.
\newblock In \emph{\bibinfo{booktitle}{Proceedings of the IEEE Computer Society
  Conference on Computer Vision and Pattern Recognition}}, vol.
  \bibinfo{volume}{2016-Decem}, \bibinfo{pages}{770--778}
  (\bibinfo{year}{2016}).
\newblock \eprint{arXiv:1512.03385v1}.

\bibitem{Szegedy2015}
\bibinfo{author}{Szegedy, C.} \emph{et~al.}
\newblock \bibinfo{title}{{Going deeper with convolutions}}.
\newblock \emph{\bibinfo{journal}{Proceedings of the IEEE Computer Society
  Conference on Computer Vision and Pattern Recognition}}
  \textbf{\bibinfo{volume}{07-12-June}}, \bibinfo{pages}{1--9}
  (\bibinfo{year}{2015}).
\newblock \eprint{arXiv:1409.4842v1}.

\bibitem{Donahue2013}
\bibinfo{author}{Donahue, J.} \emph{et~al.}
\newblock \bibinfo{title}{{DeCAF: A Deep Convolutional Activation Feature for
  Generic Visual Recognition}}.
\newblock \emph{\bibinfo{journal}{CoRR}} \textbf{\bibinfo{volume}{abs/1310.1}}
  (\bibinfo{year}{2013}).
\newblock \urlprefix\url{http://arxiv.org/abs/1310.1531}.
\newblock \eprint{1310.1531}.

\bibitem{Zeiler13}
\bibinfo{author}{Zeiler, M.~D.} \& \bibinfo{author}{Fergus, R.}
\newblock \bibinfo{title}{{Visualizing and Understanding Convolutional
  Networks}}.
\newblock \emph{\bibinfo{journal}{CoRR}} \textbf{\bibinfo{volume}{abs/1311.2}}
  (\bibinfo{year}{2013}).
\newblock \urlprefix\url{http://arxiv.org/abs/1311.2901}.
\newblock \eprint{1311.2901}.

\bibitem{coco}
\bibinfo{author}{Lin, T.-Y.} \emph{et~al.}
\newblock \bibinfo{title}{{Microsoft COCO: Common Objects in Context}}.
\newblock In \bibinfo{editor}{Fleet, D.}, \bibinfo{editor}{Pajdla, T.},
  \bibinfo{editor}{Schiele, B.} \& \bibinfo{editor}{Tuytelaars, T.} (eds.)
  \emph{\bibinfo{booktitle}{Computer Vision -- ECCV 2014}},
  \bibinfo{pages}{740--755} (\bibinfo{publisher}{Springer International
  Publishing}, \bibinfo{address}{Cham}, \bibinfo{year}{2014}).

\bibitem{Imagenet}
\bibinfo{author}{Russakovsky, O.} \emph{et~al.}
\newblock \bibinfo{title}{{ImageNet Large Scale Visual Recognition Challenge}}.
\newblock \emph{\bibinfo{journal}{International Journal of Computer Vision}}
  \textbf{\bibinfo{volume}{115}}, \bibinfo{pages}{211--252}
  (\bibinfo{year}{2015}).
\newblock \eprint{arXiv:1409.0575v3}.

\bibitem{barak2019}
\bibinfo{author}{Barak, Y.}, \bibinfo{author}{Srivastava, A.} \&
  \bibinfo{author}{Osovski, S.}
\newblock \bibinfo{title}{Correlating fracture toughness and fracture surface
  roughness via correlation length scale}.
\newblock \emph{\bibinfo{journal}{International Journal of Fracture}}
  \bibinfo{pages}{1--12} (\bibinfo{year}{2019}).

\bibitem{Thoma2016}
\bibinfo{author}{Thoma, M.}
\newblock \bibinfo{title}{{A Survey of Semantic Segmentation}}.
\newblock \emph{\bibinfo{journal}{CoRR}} \textbf{\bibinfo{volume}{abs/1602.0}},
  \bibinfo{pages}{1--16} (\bibinfo{year}{2016}).
\newblock \urlprefix\url{http://arxiv.org/abs/1602.06541}.
\newblock \eprint{1602.06541}.

\bibitem{Gould2008}
\bibinfo{author}{Gould, S.}, \bibinfo{author}{Rodgers, J.},
  \bibinfo{author}{Cohen, D.}, \bibinfo{author}{Elidan, G.} \&
  \bibinfo{author}{Koller, D.}
\newblock \bibinfo{title}{{Multi-class segmentation with relative location
  prior}}.
\newblock \emph{\bibinfo{journal}{International Journal of Computer Vision}}
  \textbf{\bibinfo{volume}{80}}, \bibinfo{pages}{300--316}
  (\bibinfo{year}{2008}).

\bibitem{Shotton2006}
\bibinfo{author}{Shotton, J.}, \bibinfo{author}{Winn, J.},
  \bibinfo{author}{Rother, C.} \& \bibinfo{author}{Criminisi, A.}
\newblock \bibinfo{title}{{TextonBoost: Joint appearance, shape and context
  modeling for multi-class object recognition and segmentation}}.
\newblock In \emph{\bibinfo{booktitle}{Lecture Notes in Computer Science
  (including subseries Lecture Notes in Artificial Intelligence and Lecture
  Notes in Bioinformatics)}}, vol. \bibinfo{volume}{3951 LNCS},
  \bibinfo{pages}{1--15} (\bibinfo{year}{2006}).

\bibitem{Nair2010}
\bibinfo{author}{Nair, V.} \& \bibinfo{author}{Hinton, G.~E.}
\newblock \bibinfo{title}{{Rectified Linear Units Improve Restricted Boltzmann
  Machines}}.
\newblock In \emph{\bibinfo{booktitle}{Proceedings of the 27th International
  Conference on International Conference on Machine Learning}},
  \bibinfo{number}{3} (\bibinfo{year}{2010}).
\newblock
  \urlprefix\url{https://www.cs.toronto.edu/{~}hinton/absps/reluICML.pdf}.
\newblock \eprint{1111.6189v1}.

\bibitem{Long2015}
\bibinfo{author}{Long, J.}, \bibinfo{author}{Shelhamer, E.} \&
  \bibinfo{author}{Darrell, T.}
\newblock \bibinfo{title}{{Fully Convolutional Networks for Semantic
  Segmentation}}.
\newblock In \emph{\bibinfo{booktitle}{The IEEE Conference on Computer Vision
  and Pattern Recognition (CVPR)}} (\bibinfo{year}{2015}).

\bibitem{Badrinarayanan2017}
\bibinfo{author}{Badrinarayanan, V.}, \bibinfo{author}{Kendall, A.} \&
  \bibinfo{author}{Cipolla, R.}
\newblock \bibinfo{title}{{SegNet: A Deep Convolutional Encoder-Decoder
  Architecture for Image Segmentation.}}
\newblock \emph{\bibinfo{journal}{IEEE transactions on pattern analysis and
  machine intelligence}} \textbf{\bibinfo{volume}{39}},
  \bibinfo{pages}{2481--2495} (\bibinfo{year}{2017}).
\newblock \urlprefix\url{http://www.ncbi.nlm.nih.gov/pubmed/28060704}.
\newblock \eprint{arXiv:1511.00561v3}.

\bibitem{Ronneberger2015}
\bibinfo{author}{Ronneberger, O.}, \bibinfo{author}{Fischer, P.} \&
  \bibinfo{author}{Brox, T.}
\newblock \bibinfo{title}{{U-net: Convolutional networks for biomedical image
  segmentation}}.
\newblock \emph{\bibinfo{journal}{Lecture Notes in Computer Science (including
  subseries Lecture Notes in Artificial Intelligence and Lecture Notes in
  Bioinformatics)}} \textbf{\bibinfo{volume}{9351}}, \bibinfo{pages}{234--241}
  (\bibinfo{year}{2015}).
\newblock \eprint{arXiv:1505.04597v1}.

\bibitem{chollet2015keras}
\bibinfo{author}{Chollet, F.} \& \bibinfo{author}{Others}.
\newblock \bibinfo{title}{{Keras}}.
\newblock \bibinfo{howpublished}{$\backslash$url{\{}https://keras.io{\}}}
  (\bibinfo{year}{2015}).

\bibitem{Abadi2016}
\bibinfo{author}{Abadi, M.} \emph{et~al.}
\newblock \bibinfo{title}{{TensorFlow: Large-Scale Machine Learning on
  Heterogeneous Distributed Systems}}.
\newblock vol. \bibinfo{volume}{abs/1603.0} (\bibinfo{year}{2016}).
\newblock \urlprefix\url{http://arxiv.org/abs/1603.04467}.
\newblock \eprint{1603.04467}.

\bibitem{tensorflow2015-whitepaper}
\bibinfo{author}{Martin, A.} \emph{et~al.}
\newblock \bibinfo{title}{{{\{}TensorFlow{\}}: Large-Scale Machine Learning on
  Heterogeneous Systems}} (\bibinfo{year}{2015}).
\newblock \urlprefix\url{https://www.tensorflow.org/}.

\bibitem{blumer}
\bibinfo{author}{Blumer, D.} \& \bibinfo{author}{Rittel, D.}
\newblock \bibinfo{title}{The influence of microstructure on the static and
  dynamic strength of transparent magnesium aluminate spinel (mgal2o4)}.
\newblock \emph{\bibinfo{journal}{Journal of the European Ceramic Society}}
  \textbf{\bibinfo{volume}{38}}, \bibinfo{pages}{3618--3634}
  (\bibinfo{year}{2018}).

\bibitem{quinn2016}
\bibinfo{author}{Quinn, G.~D.}
\newblock \emph{\bibinfo{title}{{Fractography of ceramics and glasses}}}
  (\bibinfo{publisher}{National Institute of Standards and Technology
  Washington, DC}, \bibinfo{year}{2016}).

\bibitem{dutta2019vgg}
\bibinfo{author}{Dutta, A.} \& \bibinfo{author}{Zisserman, A.}
\newblock \bibinfo{title}{{The {\{}VIA{\}} Annotation Software for Images,
  Audio and Video}}.
\newblock \emph{\bibinfo{journal}{arXiv preprint arXiv:1904.10699}}
  (\bibinfo{year}{2019}).

\bibitem{dutta2016via}
\bibinfo{author}{Dutta, A.}, \bibinfo{author}{Gupta, A.} \&
  \bibinfo{author}{Zissermann, A.}
\newblock \bibinfo{title}{{{\{}VGG{\}} Image Annotator ({\{}VIA{\}})}}.
\newblock
  \bibinfo{howpublished}{http://www.robots.ox.ac.uk/{\~{}}vgg/software/via/}
  (\bibinfo{year}{2016}).

\bibitem{Kingma2014}
\bibinfo{author}{Kingma, D.~P.} \& \bibinfo{author}{Ba, J.}
\newblock \bibinfo{title}{{Adam: A Method for Stochastic Optimization}}.
\newblock \emph{\bibinfo{journal}{CoRR}} \bibinfo{pages}{1--15}
  (\bibinfo{year}{2014}).
\newblock \urlprefix\url{http://arxiv.org/abs/1412.6980}.
\newblock \eprint{1412.6980}.

\end{thebibliography}
\bibliographystyle{naturemag}



\end{document}